\documentclass[journal]{IEEEtran}
\usepackage{color}
\usepackage{graphicx}
\DeclareGraphicsExtensions{.pdf}
\graphicspath{{./fig/}}

\usepackage[cmex10]{amsmath}
\usepackage{amssymb, latexsym}
\usepackage{amsfonts}
\usepackage{amsthm}
\usepackage{hyperref}
\hypersetup{
    colorlinks=true,
    linkcolor=blue,
    filecolor=magenta,      
    urlcolor=cyan,
}
\usepackage{caption}
\usepackage{subcaption}
\newcommand\widebar[1]{\mathop{\overline{#1}}}
\newcommand{\be}{\begin{equation}}
\newcommand{\ee}{\end{equation}}
\def\bal#1\eal{\begin{align}#1\end{align}}

\begin{document}
%\begin{frontmatter}
%\title{Audio scene monitoring using self-calibrating microphone arrays}
\title{Audio scene monitoring using redundant un-localized microphone arrays}
%There are wavelengths people cannot see,  sounds people cannot hear, and maybe computers have thoughts that people cannot think.

\author{Peter Gerstoft, \IEEEmembership{Senior Member,~IEEE,}
Yihan Hu, Chaitanya Patil,
%Yifan Wu, 
Ardel Alegre, Michael J. Bianco, \IEEEmembership{Member,~IEEE}
Yoav Freund, and
Fran\c{c}ois Grondin,  \IEEEmembership{Member,~IEEE}

\thanks{ P. Gerstoft, Y. Hu, C. Patil,  A. Alegre, M.J. Bianco 
and Y. Freund are with University of California San Diego, La Jolla, CA 92093-0238,USA, http://noiselab.ucsd.edu}
\thanks{F. Grondin is with Universit\'e de Sherbrooke, Sherbrooke, Qu\'ebec, Canada}
%\thanks{Supported by the Office of Naval Research, Grant No. N00014-18-1-2118.}
}
\maketitle

\begin{abstract}
We present a system for localizing sound sources in a room with several microphone arrays.
Unlike most existing approaches, the positions of the arrays in space are assumed to be unknown.
Each circular array performs direction of arrival (DOA) estimation independently. 
The DOAs are then fed to a fusion center where they are concatenated and used to perform the localization based on two proposed methods, which require only few labeled source locations for calibration.
The first proposed method is based on principal component analysis (PCA) of the observed DOA and does not require any calibration.
The array cluster can then perform localization on a manifold defined by the PCA of concatenated DOAs over time.
The second proposed method performs localization using an affine transformation between the DOA vectors and the room manifold. 
The PCA approach has fewer requirements on the training sequence, but is less robust to missing DOAs from one of the arrays.
The approach is demonstrated with a set of five 8-microphone circular arrays, placed at unknown fixed locations in an office. Both the PCA approach and the direct approach can easily map out a rectangle based on a few calibration points with similar accuracy as calibration points.
The methods demonstrated here provide a step towards monitoring activities in a smart home and require little installation effort as the array locations are not needed.

\today
 \end{abstract}
\IEEEpeerreviewmaketitle

\section{Introduction}
Microphone arrays, in the form of smart speakers, have become an affordable household item. As a result, these systems are ubiquitous, and there may be many microphone arrays in a single room. 
Most audio array systems require knowledge of the relative locations and orientations of the arrays. 
We instead use a few source calibration points with known relative locations, which are easier to implement. 
By using redundant arrays, we obtain higher accuracy and are less concerned with array placement.

In this paper we describe localization approaches which use several microphone arrays of unknown locations. The DOAs from each array are then collected through WiFi to a central fusion center where they are  concatenated to form a global DOA vector. Based on the global DOA vector we perform localization in a room with two proposed methods: a subspace method based on PCA and an affine mapping-based approach.

We first consider localization using principal component analysis (PCA) to obtain a low-dimensional mapping of the DOA vector observations over time and space. This PCA approach does not require knowledge of calibration source location, just an association between PCA coordinates and room coordinates. The mapping requires the relative location of calibration points, say corner of a table. While this approach does not provide physical dimensions, it can give relations between locations in a room.

We also consider a localization approach based on affine mapping between the DOA vector and room coordinates. This mapping is more robust to missing array DOAs than the PCA approach, as it is based on a physical mapping to room coordinates.
To train the mapping it also requires relative location of calibration points.

% Using  PCA, it is demonstrated that a  unique room location  gives unique PCA coordinates. Thus, source locations are clearly defined in PCA space. 

For our experiments, we have five arrays, placed at fixed unknown locations in a reverberant office environment. The arrays are connected to a Raspberry Pi 3, which runs  the Open embeddeD Audition System (ODAS) software, \cite{grondin2019lightweight,grondin2013manyears} and can track the directions of arrival (DOAs) for up to four sound sources simultaneously. This open-source framework is appealing as it allows on-board DOA estimation for each microphone array. We use this existing system to observe the sound sources and compute their DOA relative to the array.
%and produce a reconstruction of the sound for each source.

Such a setup could be useful for improving monitoring of sound events in a smart home, using existing arrays \cite{debes2016monitoring}. Since the DOA processing is performed on-board, this helps preserve privacy, as audio in this approach is not share over the network.
This could be used for fall monitoring \cite{fleury2009svm} and daily events classification \cite{virtanen2018computational}. This system could also be combined to sound scene classification, e.g., \cite{serizel2020sound,chan2020comprehensive}.  

% As shown here, having a redundant set of arrays enables tracking events in much larger details than with just a few sensors.

\subsection{Related work}

Alternative methods would combine the whole array stream of data at the { fusion center} by using 
maximum likelihood beamforming for multiple arrays \cite{Zhang2008,aarabi2003fusion} and localize the arrays in an {\it ad hoc} network \cite{gaubitch2013auto}. In our approach, where the DOAs are processed at the individual arrays, we benefit from transmitting just the DOA stream. This requires less bandwidth than using the raw array output. In terms of privacy this is a huge improvement as no signals are transmitted to the  fusion center.
In a fusion center multiple DOAs can be used for localization based on known array positions\cite{cobos2017survey}. %Our approach is not sensitive to multi-path propagation.

Resolving closely spaced acoustic sources when the noise power is varying in space and time has been studied extensively.
As part of ODAS, we rely on the low-complexity Steered-Response Power Phase Transform (SRP-PHAT) to estimate DOAs \cite{brandstein1997robust,grondin2019lightweight}.
Many other methods could be used such as adaptive beamformers \cite{VanTreesBook} or compressive DOA estimation 
\cite{XenakiCS:2014,Wipf2007, Gerstoft2015,Gerstoft2018CS}.
%using multiple measurement vectors (MMV or multiple snapshots). We solve the MMV problem using 
Sparse Bayesian learning (SBL)~\cite{Wipf2007,gerstoft2016mmv,nannuru2019,Gerstoft2019,Gemba2019} could also be used.
%For details see Ref \cite{Gerstoft2019} and it has been demonstrated on real data \cite{Gemba2019}. 
 Multiple Signal Classification (MUSIC) \cite{schmidt1986multiple} is also common for localization in noisy environments \cite{nakamura2011intelligent,nakamura2012real}.
DOA estimation could also rely on machine leaning \cite{ozanich2020feedforward,barthelme2020machine}. 
Since each circular array has well-defined geometry classical array processing seems more appropriate, and machine learning could be used in the combination of the arrays.
More features than just DOA could be beneficial for localization \cite{meyer2018scalable},
 e.g., echoes \cite{di2019mirage}.
As a low-complexity approach, just three microphones have been used \cite{rascon2015lightweight}. 

In the future, more advanced machine learning (ML)\cite{Bianco2019} could help with this task as it already did in related localization problems. In particular, ML approaches might better-address the inherent non-linearity in sound localization.
Neural network classifier \cite{chakrabarty2019multi,adavanne2019sound,ozanich2020feedforward,wu2020}
or semi supervised learning\cite{bianco2020,bianco2021access,Hu2020} would be of first interest.
%
%There has been great interest in machine learning (ML)-based techniques  in acoustics, including source localization and event detection \cite{nakashima20053d,deleforge20122d,mesaros2017dcase,vincent2018audio,ping2020three,zhu2020feature}.

Many indoor localization systems have been proposed \cite{liu2020indoor}. Sound source localization (SSL) also has widespread applications in human-robot interaction \cite{nakadai2000active}, ocean acoustics \cite{gemba2019robust}, teleconferencing \cite{zhao2012real}, drone localization \cite{lauzon2017localization}, and automatic speech recognition \cite{xiao2014ntu}. These have been designed for various applications as indoor navigation, communication and health.
In general, they involve a first step where receiver geometric configuration such as distances and angles are measured. In the second step, the target is located using the measured data. Our  proposed system replaces the first step with a training/calibration step.

DOAs are the only features used here. Other systems also use travel time difference of arrival between the arrays \cite{dang2021}. This requires a precise relative clock for each array system, maybe requiring more hardware. The actual system implemented here cost about 200 USD for one array.
The proposed system is demonstrated with active sources at well-known points. This serves to better validate the approach.

\begin{figure}[t]
    \centering
    \includegraphics[width=0.7\linewidth]{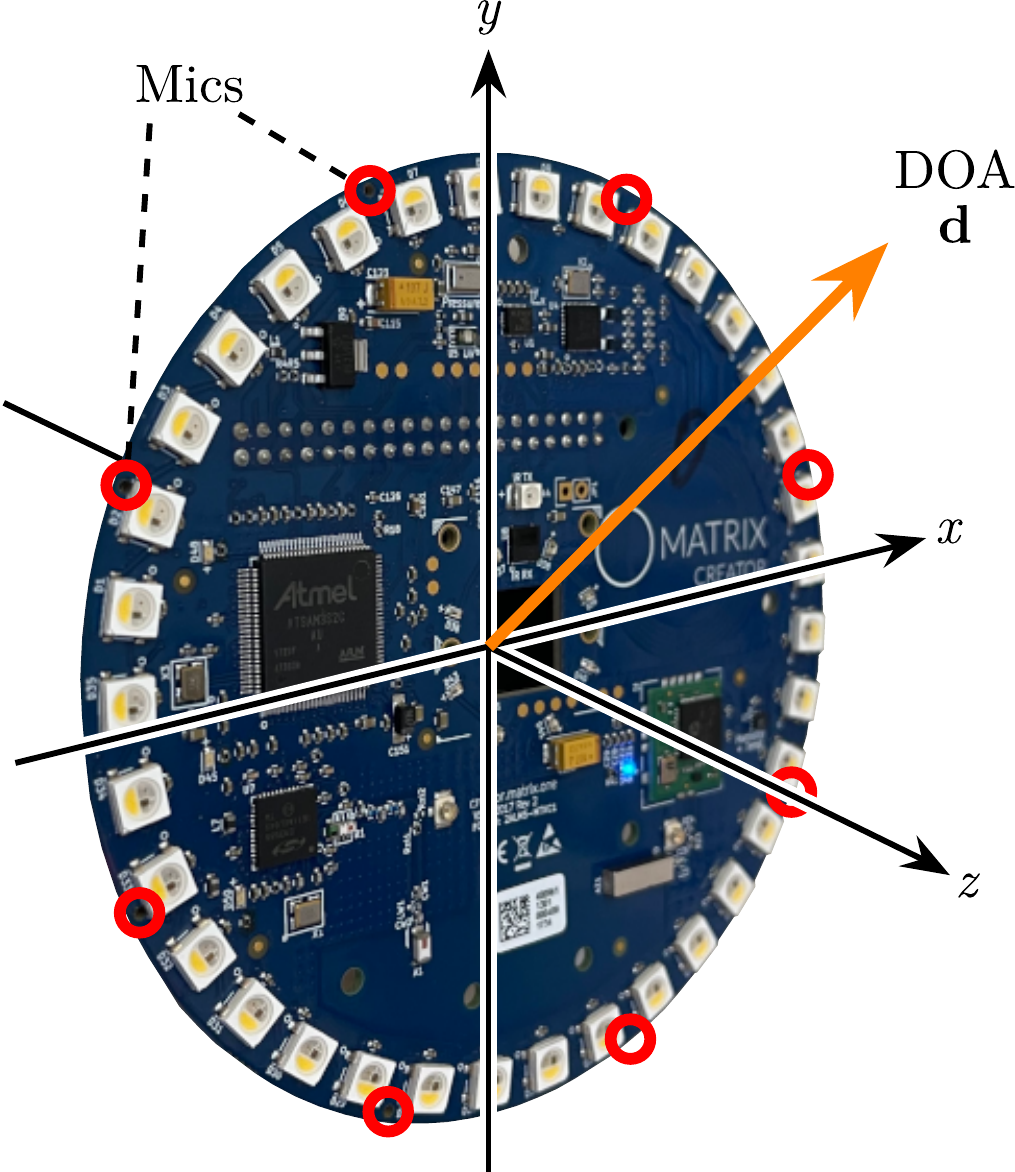}
    \caption{The Matrix Creator array is equipped with 8 microphones (circled in red), disposed on a 10 cm diameter circle. A unit vector in the positive $z$-plane points in the direction of the DOA (in orange, denoted by $\mathbf{d}$).}
    \label{fig:arraysetup}
\end{figure}

%%%%%%%%%%%%%%%%%
\section{Processing localization}
We assume a room equipped with $M$ time-synchronized circular microphone arrays, which provides the DOAs based on a local coordinate system.
The DOA for each array lies on a virtual unit half-sphere in the positive $z$-plane (see Fig. \ref{fig:arraysetup}), and is defined as ${\bf d}= [d_x \ d_y \ d_z ]^T$, where $d_x^2+d_y^2+ d_z^2=1$ and $d_z \ge 0$. Our development and processing use the output DOA vectors of the time-synchronized arrays as they arrive at the fusion center.

From the M arrays we obtain one $3M$ dimensional DOA vector ${\bf d}=[{\bf d}_1^T \dots  {\bf d}_M^T ]^T $ from one sound source. The array positions are unknown. Thus, to perform source localization we first need to do training  based on previous $L$ observations of the ${\bf d}$ vector, collected into a ${3M\times L}$ matrix $\bf D$, called calibration DOAs. We propose two methods:
\\
%\begin{enumerate}
 %   \item[
    {\bf Sec.~\ref{se:pca}} method relies on observing  $\bf D$ and then performing principal component analysis (PCA) on this matrix. Thus, we localize the source in a PCA space. The location of the sources generating  the  $\bf D$ matrix can be unknown.
    \\
    {\bf Sec.~\ref{se:lin}--\ref{se:disturb}} approach assumes an affine transformation between  ${3M}$ dimensional ${\bf d}$-vector and the local $N=3$ dimensional room.
%\end{enumerate}

For the first approach  we need all (or the same) arrays observing the sound source of interest, but no calibration sources are needed.
For the second approach  we need a set of calibration sources (more than 3).

%%%%%%%%%%%%%%%%%%%
\subsection{PCA}\label{se:pca}
Performing PCA on the observed DOA vectors gives ${\bf D}= {\bf U}{\bf \Lambda }{\bf V}^T$, with ${\bf U}, {\bf V}$ the left and right singular vectors and ${\bf \Lambda }$ the singular values. Assuming the relation between DOA and spatial location is approximately linear  \eqref{eq:lin} with N (2 or 3) spatial coordinates, the $J$ first (2 or 3) singular DOA vectors should be sufficient to describe the location. Throughout the rest of the paper, we use $J=2$ first components, as all sources are mainly in the 2-D horizontal plane.

Decomposing the $i$th DOA observation ${\bf d}_i$ with only the first $J=2$ singular vectors used, the reduced  ${\bf U}_J=[{\bf u}_1 \,{\bf u}_2]$.
\begin{equation}
    {\bf d}_i\approx\sum_{j=1}^J a_{ij} {\bf u}_j={\bf U}_J{\bf a}_i, \qquad a_{ij}= {\bf d}_i^T{\bf u}_j~, \label{eq:pca}
\end{equation}
where $ {\bf a}_i=[a_{i1}\, a_{i2}]^T$ are the coefficients of the two singular vectors that define the DOA vector ${\bf d}_i$.
%For $3M=J$, the decomposition \eqref{eq:pca} is unique. 
Since physically there should only be 2--3 large components, finding the decomposition in PCA space defines an unknown room position.

In the above processing, it is assumed that all arrays observe all the sources. However, in a real system there will always be missing observations due to malfunctioning arrays, nature of the room setup, source directivity or weak sources.
This can happen in either the calibration or the mapping part of the experiments.

For mapping to PCA coordinates, omitting the missing array is easy, but the sound source will then map to a different location in PCA space. Thus, for sources evolving sequentially in time it is preferred  to only use the arrays that is continuously tracking the source.

\subsection{Affine mapping to room manifold}\label{se:lin}
In general, the mapping between  $N$-dimensional room observations $\mathbf{r}\in\mathbb{R}^N$ ($N$ is 2 or 3) and $3M$-dimensional DOA vector $\mathbf{d}$ is non-linear and unknown 
\begin{equation}
{\bf r}= {\bf f}({\bf d}). 
\end{equation}
The non-linear variation of the DOA vector $\bf d$ vs spatial location for one array is indicated in Fig. \ref{fig:odas}.
We assume a linear mapping (affine transformation)  might be sufficient for this mapping of a room. 
%This is more likely to work well if each component of the DOA $\bf d$ only have minor variations and are thus nearly linear. However, 
This was observed to work well for the DOA variation in the room used here, see Sec.~\ref{sec:exp}. The affine transformation is 
\begin{equation}
    {\bf r}= {\bf r}_0+ {\bf B}{\bf d} \label{eq:lin}
\end{equation}
where ${\bf r}_0\in\mathbb{R}^N$ is the offset and ${\bf B}\in\mathbb{R}^{N\times 3M}$ are the linear coefficients. Both  ${\bf r}_0$ and ${\bf B}$ are determined in Sec.~\ref{se:calibration} by performing recording at $K$ locations ${\bf r}_k$.

%\subsection{Missing array observations}
Typically, a weak source is not observed on all arrays, giving a DOA observation $\bf d$ with non-active elements, we then retain the active elements in ${\bf d}_a$ produced by the list of active arrays $I_a$. 
From the calibration DOA matrix $\bf D$ we only use the active arrays $I_a$.
For each DOA observation ${\bf d}$ and $I_a$, we then determine the ${\bf \sf B}(I_a)$ function  with only the entries of the calibration DOA $\bf D$ corresponding to active arrays $I_a$ and use this to determine the $\bf B$ matrix for just $I_a$. This $\bf B$ matrix is then multiplied with the active DOAs ${\bf d}_a$. The non-active elements are not used in the mapping. This gives
\begin{equation}
    {\bf r}= {\bf r}_0+ {\bf \sf B}(I_a)\,{\bf d}_a ~.\label{eq:Nonlin}
\end{equation}
For each case of active DOAs, the $\bf B$ matrix is determined as in Sec.~\ref{se:calibration}.

%For the mapping to room coordinates \eqref{eq:lin}, the easiest solution is to discard the array entries with missing observations for both calibration and mapping, recalculate $\bf B$-matrix and perform mapping  \eqref{eq:Nonlin}. Since the true relation between DOA and room coordinates is nonlinear, the recalculated $\bf B$-matrix is not just a scaling of the original.

%%%%%%%%%%%%%%%%%%%
\subsection{Linear mapping calibration}\label{se:calibration}
We perform a sound source recording at $K$ locations, for location ${\bf r}_k$ we have  $L_k$ observations ${\bf D}_k =[{\bf d}_1\dots {\bf d}_{L_k}] $.
%\in [ 0\,\, 1]^{N \times {L_k}}$.
For all $L=\sum_{k=1}^K L_k$ observations, we obtain the ${3M\times L}$ matrix of DOA observations ${\bf D}=[{\bf D}_1\, \dots {\bf D}_K]$
generated from a source with corresponding
 location matrix 
\begin{equation}
{\bf R} =[{\bf r}_1 {\bf i}^T_{L_1}\dots {\bf r}_K {\bf i}^T_{L_K}] \in \mathbb{R}^{N\times L}
\end{equation}
where ${\bf i}_{L_j}$ is a vector of length $L_j$ with all ones.

Defining the cost function (using the trance operator $\mbox{Tr}$ and Frobenius norm $\|\cdot\|_f $)
\begin{eqnarray}
\phi&=&\|{\bf R} -({\bf r}_0 {\bf i}^T_{L}+{\bf B}{\bf D})\|_F^2= \mbox{Tr}[{\bf E}^T{\bf E}]
\nonumber \\
{\bf E}&=&{\bf R} -({\bf r}_0 {\bf i}^T_{L}+{\bf B}{\bf D})~.
\end{eqnarray}
Differentiating the cost function
\begin{align}
\frac{\partial \phi }{\partial {\bf E}}&= 2 {\bf E}
\\
\frac{\partial \phi }{\partial {\bf r}_0}&=-2[{\bf R} -({\bf r}_0 {\bf i}^T_{L}+{\bf B}{\bf D})]{\bf i}_{L} \label{eq:rd}
\\
\frac{\partial \phi }{\partial {\bf B}}&=-2{\bf E}{\bf D}^T~.\label{eq:bd}
\end{align}

Forcing \eqref{eq:rd} to zero gives ( $\bar{\cdot}$ is the mean operator over observations $L$)
\begin{equation}
{\bf r}_0=\frac{1}{L}[{\bf R} -{\bf B} {\bf D}]{\bf i}_L
=\widebar{{\bf r}-{\bf B} {\bf d}} 
=\widebar{\bf r}-{\bf B} \widebar{\bf d}
\end{equation}
Forcing \eqref{eq:bd} to zero gives
\begin{align}
{\bf E}{\bf D}^T&={\bf 0}
\nonumber \\
{\bf B}{\bf D}{\bf D}^T&=[{\bf R} -{\bf r_0} {\bf i}^T_{L}]{\bf D}^T
\nonumber \\
{\bf B}{\bf D}{\bf D}^T&=[{\bf R} -(\widebar{\bf r}-{\bf B} \widebar{\bf d} ){\bf i}^T_{L}]{\bf D}^T
\nonumber \\
{\bf B}({\bf D}-\widebar{\bf d}{\bf i}^T_{L}){\bf D}^T
&=[{\bf R} -\widebar{\bf r} {\bf i}^T_{L}]{\bf D}^T
\nonumber \\
{\bf B}&=[{\bf R} -\widebar{\bf r} {\bf i}^T_{L}]{\bf D}^T
[({\bf D}-\widebar{\bf d}{\bf i}^T_{L}){\bf D}^T]^{+}  ~,       \label{eq:bmat}
\end{align}

where $^+$ denotes the Moore-Penrose pseudo-inverse. Thus, the mapping is determined from \eqref{eq:bmat}. The number of calibration points needed depends on the uniqueness of the points in spatial and DOA space. In general, for an $N$-dimensional room space, $N+1$ calibration points spanning the space are sufficient. Using just 2 calibration points maps all observations to a line between the points and 3 calibration point is sufficient spanning a 2D plane.

%%%%%%%%%%%%%%
\subsection{Disturbances from reference point}\label{se:disturb}
For a known reference  point between a DOA vector and room location $({\bf d}_i, {\bf r}_i)$, an approximate mapping from  a measurement ${\bf d}_n$ to a new room location ${\bf r}_n$ is for a given $\bf B$ matrix:
\begin{eqnarray}
{\bf r}_n-{\bf r}_i= \frac{\partial {\bf r} }{\partial {\bf d}}\Delta {\bf d}
= {\bf B}({\bf d}_n-{\bf d}_i) ~. \label{eq:small}
\end{eqnarray}
For a given $\bf B$ matrix learned though the calibration points, we do the mapping by selecting a reference point $({\bf d}_i, {\bf r}_i)$ close to the observed ${\bf d}_n$ and use  \eqref{eq:small}.

 Similarly, for a known  DOA PCA components $ {\bf a}_i$ (as in \eqref{eq:pca}) and room location  ${\bf r}_i$ and given a PCA observation  $ {\bf a}_n $  the new room location ${\bf r}_n$ is
 \begin{equation}
     {\bf r}_n- {\bf r}_i=  {\bf B}({\bf d}_n-{\bf d}_i)= {\bf B}{\bf U} ({\bf a}_n-{\bf a}_i)=  {\bf C} ({\bf a}_n-{\bf a}_i) ~, \label{eq:pca-to-r}
 \end{equation}
  and ${\bf C}={\bf B}{\bf U}$ describes the mapping from the DOA PCA components  to spatial location. From \eqref{eq:pca-to-r} it is clear that there are an approximately a linear relation between PCA space and  location.

\subsection{Unknown calibration source location}
The PCA method comes handy when it is impractical to measure the exact positions of sound sources for calibration.
Thus, we perform the $K$ source experiments with unknown locations and compute the singular vectors from these observations to enable PCA.
Then for a new source we perform localization in this PCA space and describe  the solution in term of proximity to the $K$ sources in PCA space.

\begin{figure}[t] % FIGURE 2
    \centering
    \includegraphics[width=\linewidth]{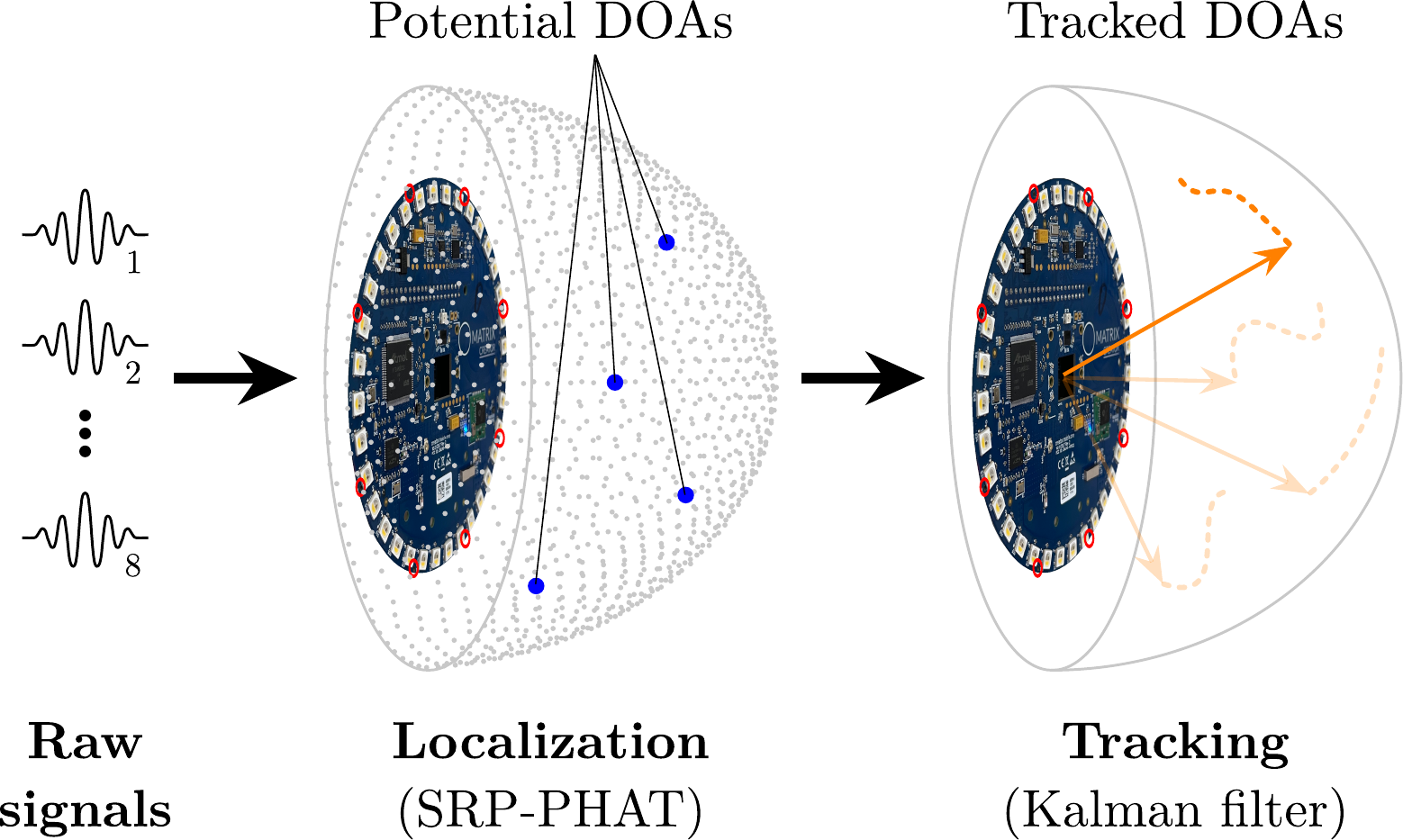}
    \caption{For each frame in time, ODAS selects four potential DOAs (out of the 1321 points) corresponding to the maximum power of SRP-PHAT, and then filters each DOA with a Kalman filter. ODAS can track up to four simultaneous sources, but in this work only the loudest source is extracted.}
    \label{fig:odas}
\end{figure}

\begin{figure}[t] % FIGURE 3

\centering
{\large a)}\\
\includegraphics[width=\columnwidth]{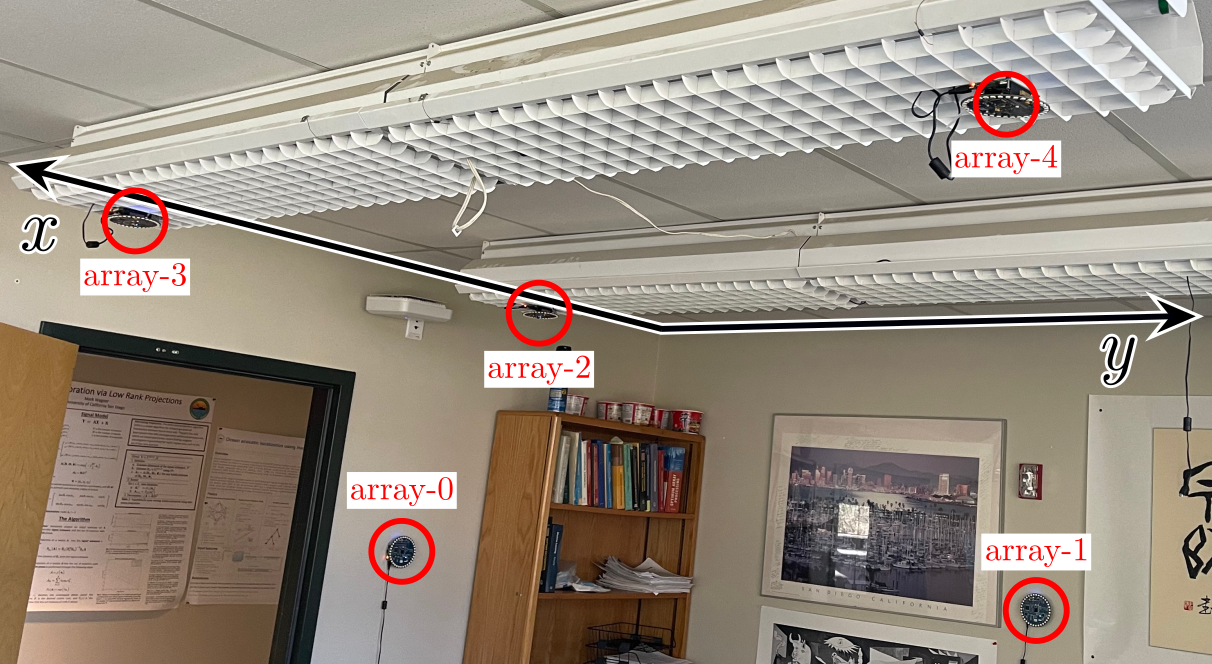}
	\vspace{5pt}
{\large b)}\\	
	\includegraphics[width=\columnwidth]{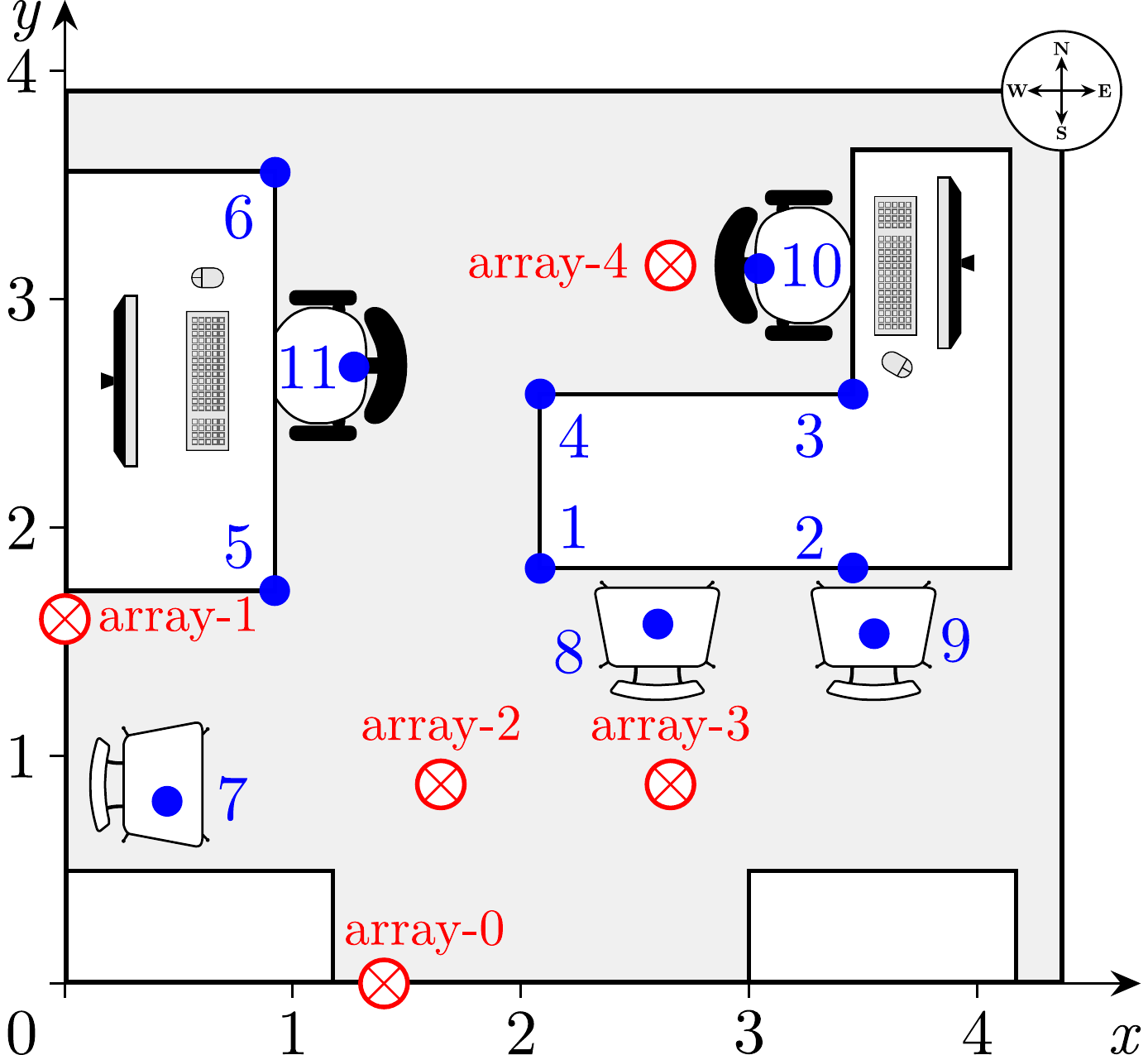}
\caption{(a) Photo of the five ODAS-arrays mounted in ceiling lamps (array-2, array-3, array-4)  and on walls (array-0, array-1). (b) Layout of room with the five arrays (red $\otimes$), the 11 calibration speaker points (blue $\bullet$) 
and coordinate system with north and $y$ direction aligned.}
\label{fig:room}
\end{figure}

%%%%%%%%%%%%%%%%
\section{Experimental design}

\subsection{Circular microphone array}
We used arrays that are part of development kit called Matrix Creator\footnote{\url{https://www.matrix.one/products/creator}} with software (ODAS\footnote{\url{http://odas.io}}) that computes the DOA using the Steered response power with phase transform (SRP-PHAT)\cite{brandstein1997robust,grondin2019lightweight} as summarized below as background information.

Each array consists of 8 microphones, uniformly placed in circle with 10 cm diameter, and connected to a Raspberry Pi 3 device\footnote{\url{https://www.raspberrypi.org/products/raspberry-pi-3-model-b/}}.
Each device runs its own instance of the ODAS framework, which outputs one DOA with a 3D unit vector in the array's local coordinate system, see Fig. \ref{fig:arraysetup}.

To represent potential DOAs, the unit half sphere is discretized using 1321 points obtained by subdividing a 20-sided convex polyhedron \cite{grondin2019lightweight}, giving a solid angle of approximately $3^{\circ}$ between grid points.
A set of time difference of arrival (TDOAs) matches each DOA based on the microphone array geometry and speed of sound (assuming plane waves).
For each hypothetical DOA  $i$, ODAS computes SRP-PHAT, with power $E_i$\cite{brandstein1997robust}:
\begin{equation}
    E_i = \sum_{p=1}^{7} \sum_{q=p+1}^{8} \frac{1}{N} \sum_{k=0}^{N-1}{\frac{X_p[k] X^*_q[k]}{|X_p[k]||X_q[k]|} \exp{\left(j\frac{2\pi \tau_{i,p,q} k}{N}\right)}},
    \label{eq:srpphat}
\end{equation}
where the superscript $^*$ denotes the complex conjugate and $X_p[k]$ is the short-time Fourier transform (STFT) frame,  $p$ is the microphone index and $k$ the frequency bin index.
Each frame consists of $N$ samples, and the expression $\tau_{i,p,q}$ is the TDOA at point $i$ between  microphones $p$ and $q$.
SRP-PHAT benefits from the low-complexity of the Fast Fourier Transform (FFT) and makes sound source localization robust to indoor reverberation.
At each time step, ODAS returns up to four potential DOAs  corresponding to the points with maximum power, as shown in \eqref{eq:srpphat}.

ODAS\cite{grondin2019lightweight} relies on a tracking module to improve the stability and resolution of DOA estimation.
Speech sources include silence periods, which should be accounted for while tracking sound sources over time.
Robust detection and tracking therefore becomes an important feature.
ODAS provides the DOA of the four sources with highest SRP-PHAT power \eqref{eq:srpphat}, and then use Kalman filters to track the DOAs over time.
For this application, we restrict ODAS to track one source.

The multi-channel raw audio is sampled at 44,100 samples/s from the Matrix Creator array, resampled by ODAS at 16,000 samples/s, which then returns an updated DOA estimation with an  8 ms time resolution to the fusion center.
Figure \ref{fig:odas} shows the ODAS pipeline for the tracked DOA.

\subsection{Array setup}
An office room was equipped with $M=5$ arrays, see Fig.~\ref{fig:room}a for locations of the three arrays mounted  in ceiling and the two on the wall, and Fig.~\ref{fig:room}b for general layout of the room  with arrays and calibration points. For convenience, we ordered the arrays with the local  $y$-component pointing up (for array-0, array-1) or in the global $y$-direction (North) for array-2, array-3, and array-5).
The array locations and headings are assumed unknown. The DOA for the $M$ arrays are concatenated into one $3M$ dimensional vector ${\bf d}=[{\bf d}_1^T \dots  {\bf d}_M^T ]^T $.

\subsection{Processing chain}
Each array processes the data independently on a Raspberry Pi 3, which is synced to a common clock. The Raspberry Pi then sends the DOAs to a fusion center, where all DOAs arrays are stored in a database after careful time alignment. All data in the next sections were extracted from this database.

\begin{figure} % FIGURE 4
\centering
	\includegraphics[width=1\columnwidth]{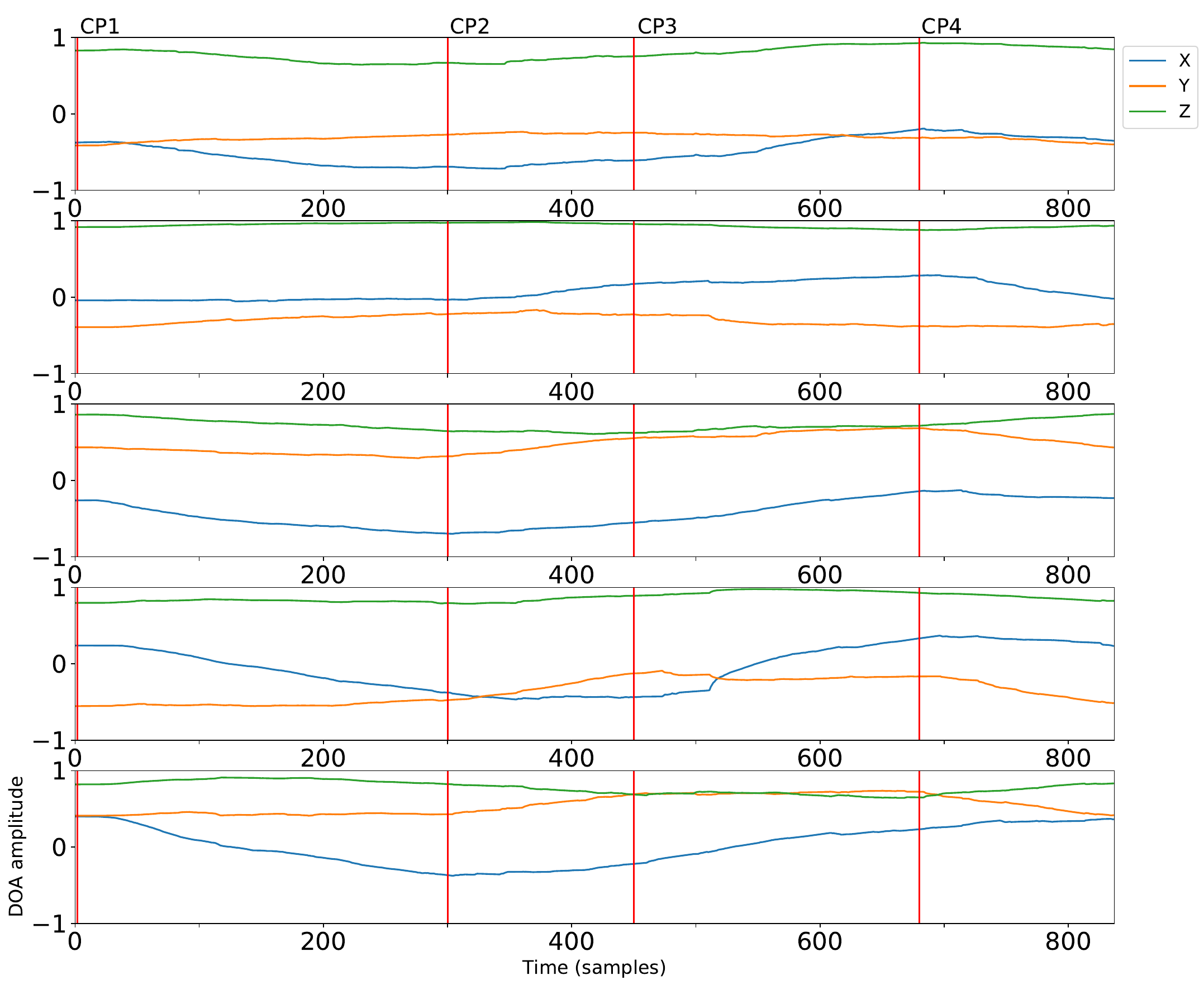}
\caption{DOAs measured from the five arrays (array-0 though array-4), here shown for measurement along the table-edge between the four calibration-points CP 1-2-3-4 (vertical line).  }
\label{fig:DOA}
\end{figure}

\begin{figure} % FIGURE 5
\centering
	\includegraphics[width=0.9\columnwidth]{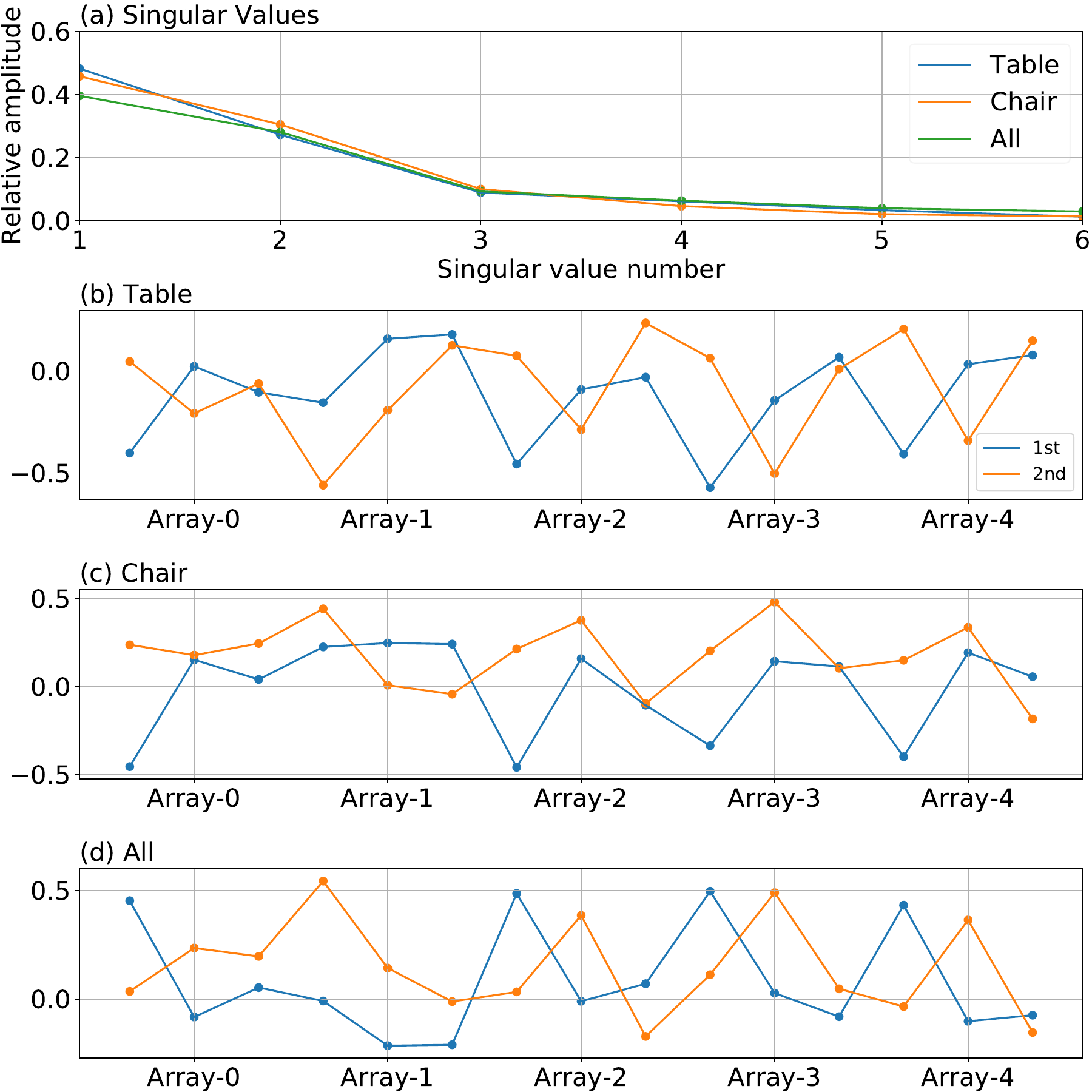}
\caption{Based on the calibration matrix $\bf D$ with 15 rows a) Normalized singular values (amplitudes as a fraction of sum of amplitudes) and  First $J$=2  DOA left singular vectors with 15 elements based on points from (b) table (calibration points 1--6), (c) chairs (calibration points 7--11), and (d) chair+table (calibration points 1--11).   
On the horizontal axis, each array shows 3 points corresponding to  the $x, y ,z$ component. Apart from sign change, the singular vectors appear similar. Note the effect of array orientation on sub-vectors of singular vector.}
\label{fig:svd}
\end{figure}

\begin{figure} % FIGURE 5
\centering
	\includegraphics[width=\columnwidth]{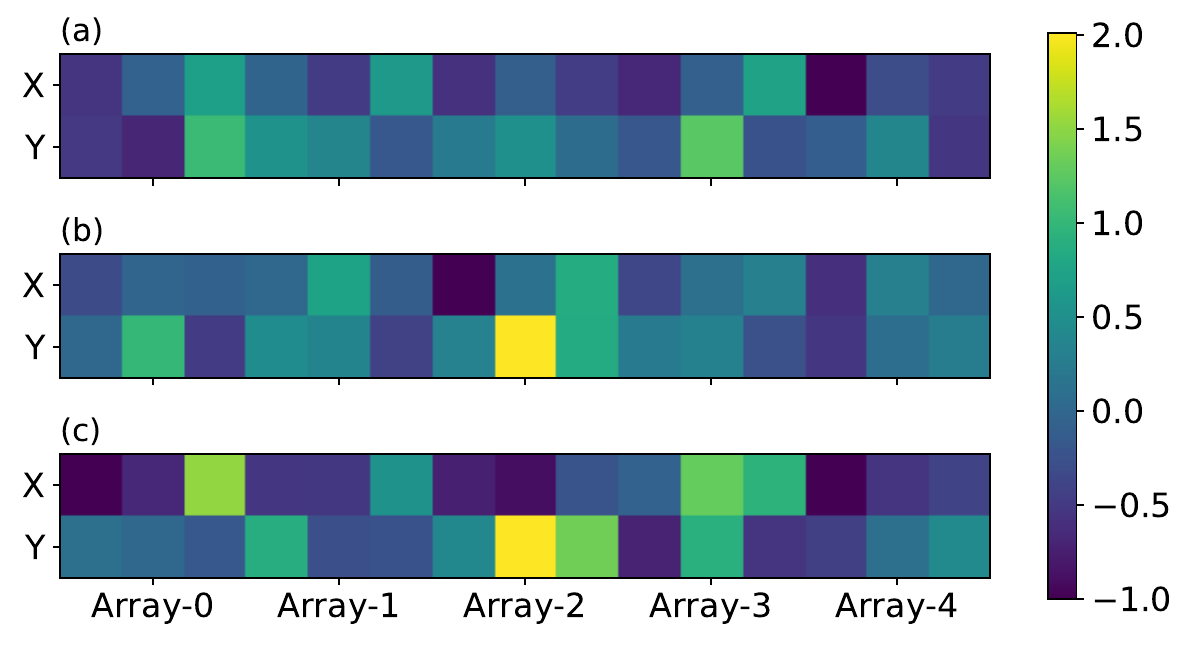}
\caption{Estimated $\bf B$ matrix \eqref{eq:bmat} based on (a) table (calibration points 1--6), (b) chair (calibration points 7--11), and (c) chair+table (calibration points 1--11).  On the horizontal axis, each array shows 3 points corresponding to the $x, y ,z$ component.  }
\label{fig:bmatrix}
\end{figure}

\begin{figure*} % FIGURE 6
\centering
	\includegraphics[width=0.95\textwidth]{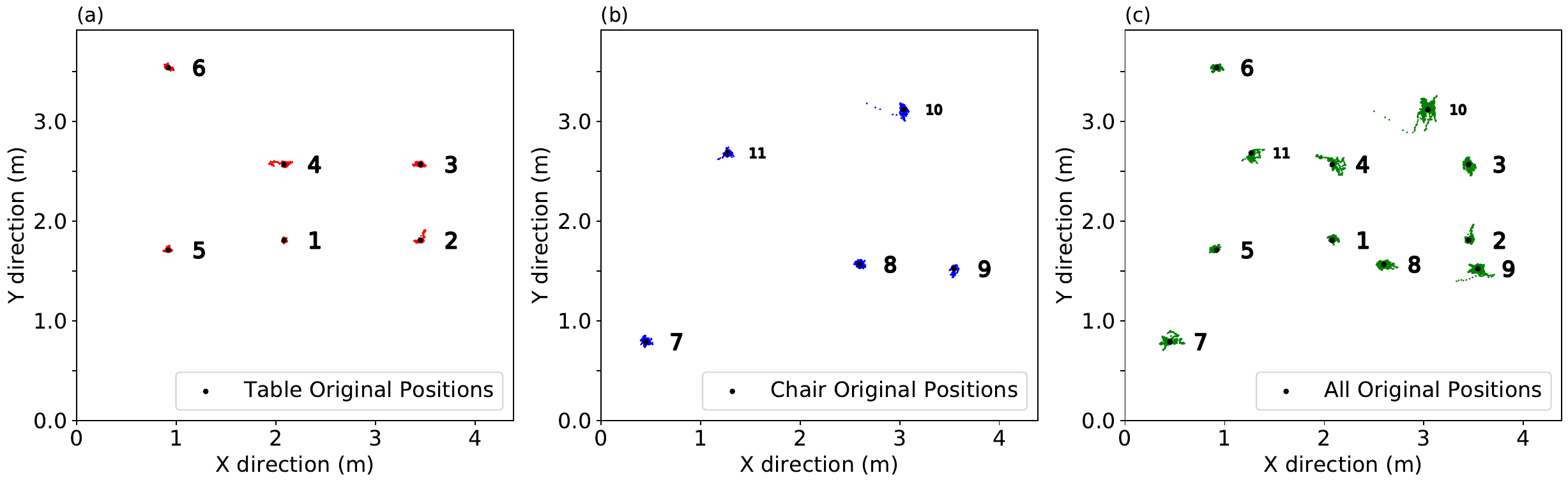}
\caption{Affine mapping \eqref{eq:lin} using estimated $\bf B$ matrix from calibration points on (a)  table (point 1--6), (b) chair (point 7--11), and (c) chair+table (point 1--11).   }
\label{fig:map}
\end{figure*}

\section{Experimental demonstration}\label{sec:exp}
The approach consists of a calibration and a mapping step.
In the calibration step, either the $\bf B$ matrix is created based on sounds from known calibration locations, or the SVD is determined using available DOAs with no need for knowing the source locations.
In the mapping step we used the observed DOA vector to map to either room coordinates \eqref{eq:lin} or PCA coordinates \eqref{eq:pca}.

\subsection{Setup}
In order to maximally activate DOA localization on each array, an omnidirectional loudspeaker (10 cm long) playing a female voice is used. 
%Measurements were performed during covid-19 lockdown, thus quiet. 
The reverberation time in the office was measured as RT60=0.3~s.
ODAS can track up to four simultaneous sound sources, but we limit it to one in these experiments.
The ODAS sound source tracking module is also tuned to track both static and moving sources.

An office room is equipped with $M=5$ arrays (see Fig.\ \ref{fig:room} for locations).
Three arrays are mounted on the ceiling (array-2, array-3, and array-4), array-0 is placed on south facing wall and array-1 is installed on west facing wall.
Note that the processing is independent on knowing array locations or orientations. 
For better physical understanding, we order the arrays with the local $y$-component pointing up for array-0, array-1, and North for array-2, array-3, and array-4.

The global $x$-axis (East) corresponds to the negative local $x$-axis for (array-0, array-2, array-3, and array-4) and $z$-axis for array-1. The global $y$-axis (North) corresponds to the positive local $y$-axis for (array-1, array-2, array-3, and array-4) and $z$-axis for array-0. These relations are for interpretation of results only, and are not used for mapping.

Each Matrix Creator array is connected to a Raspberry Pi 3 (RP3) single-board computer which provide a DOA estimate every 8 ms. The local clock of each RP3 is synchronized to a local time server at ntp.ucsd.edu over office WIFI network using the Network Time Protocol (NTP) \cite{mills1989accuracy}. When left running continuously, each local NTP client on RP3 was observed to maintain synchronization within 10 ms, but most much less. Each array generates a time-stamped digital summary of detected DOAs once every 64 ms. These Summaries are stored on local memory of the array and are pushed to the cloud every five minutes.

Every five minutes new data is down-loaded from the cloud to our { fusion center} and loaded into a relational database. The data is then sorted and joined based on the time stamp. The result is a chronological table which contains the DOAs from all arrays synchronized using the timestamps.
This table can be queried to get the full information for any period of time.

ODAS tracks the loudest sound source and transfers the DOAs to the fusion center.  At the { fusion center}, we bin the DOAs using a 64 ms time window. 
The average of all DOAs from each array is used in this 64 ms window and stored in the database. The 64 ms  resolution is sufficient as the sound sources move  slowly. Fig.~\ref{fig:DOA} shows the handheld loudspeaker being moved manually along the table edge.

We use 11 calibration points, see Fig.~\ref{fig:room}, 6 on table corners (points 1--6) and 5 on chairs (points 7--11) as shown in Fig.\ \ref{fig:room} placed at heights, 78 cm (table) and 50 cm (chair).

\subsection{Calibration}
The loudspeaker is placed at each calibration point for 30 s. For these points we record a 3-component DOA from each of the 5 arrays, giving a $3\times M=15$ element DOA-vector $\bf d$ every 0.064 s (about 450 for each point). Based on these records in $\bf D$, we extract singular vectors corresponding to measurements points on chair, table and chair+table, see Fig.~\ref{fig:svd}. Since all measurements were performed in nearly a horizontal plane, the first two PCA components carry most of the energy, see Fig.~\ref{fig:svd}a.
Focusing on  chair+table in Fig.~\ref{fig:svd}d, the first singular vector has large amplitudes, 0.5,  for $x$-component of (array-0 array-2, array-3, and array-4), all pointing westward. Thus, an increase in 1st PCA component correspond to a more westerly source.
The second singular vector has large  amplitudes, 0.5,  for $y$-component of (array-2, array-3, and array-4) and $x$-component of array-1.
An increase in the 2nd PCA component correspond to a more northerly source.

From these point measurements, the  $\bf B$-matrix \eqref{eq:bmat} is extracted  Fig.~\ref{fig:bmatrix}, we split them into batches of chair, table, and chair+table.  For all 3 mappings the $y$-component of array-2 is strongest for bottom row, corresponding to the y-component of the room. 
This indicates that an increase in y-component for array-2 gives a larger global y-component. 
Focusing on  chair+table in Fig.~\ref{fig:bmatrix}c a larger y-component of array-2 gives a larger global y-component.

Although the true mapping is non-linear, it is possible to find a linear mapping for the 5 points (Fig.~\ref{fig:map}a, table),  6 points (Fig.~\ref{fig:map}a, chair), and 11  points (Fig.~\ref{fig:map}a). For each point, ~450 DOA vectors were recorded with small fluctuations around the true DOA values. The fluctuations are from the Kalman filter output of the fixed sources. These fluctuations cause a  cloud around each point when applying the estimated $\bf B$ matrix.  The noise cloud near each point increases as more calibration points are used in Fig.~\ref{fig:map}c. This is because as number of points increases, due to non-linearity the mapping on these points become noisier.

\begin{figure} % FIGURE 7
\centering
	\includegraphics[width=0.9\columnwidth]{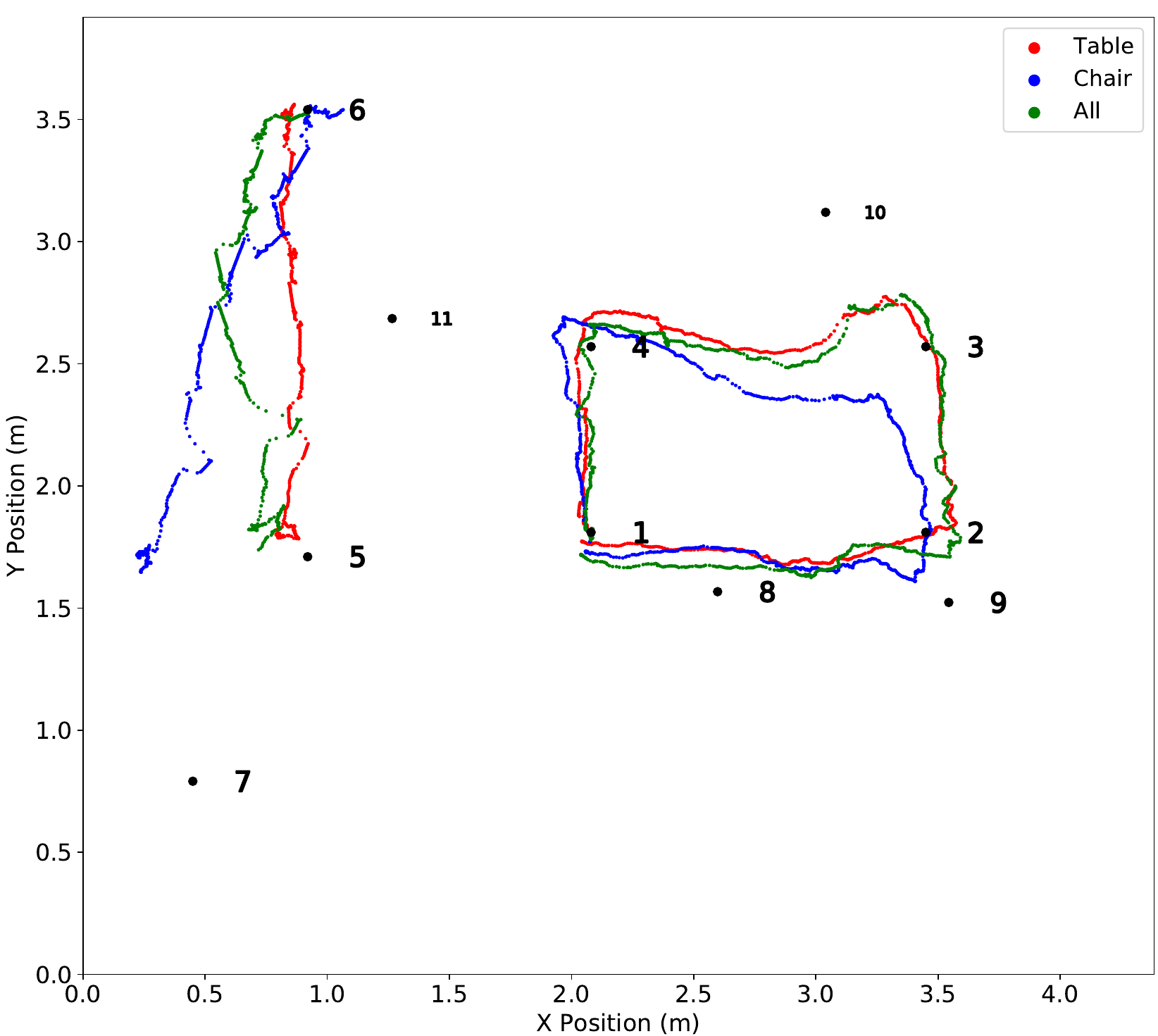}
\caption{Affine mapping \eqref{eq:lin} based on sliding source along  table edge (points 1-2-3-4) using point 1 as reference and (b) long-table edge (points 5-6) using point 6 as reference. The  estimated $\bf B$ matrix is based on  table (point 1--6), chair (point 7--11), and chair+table (point 1--11).    }
\label{fig:maptable}
\end{figure}

\begin{figure*} % FIGURE 8
\centering
	\includegraphics[width=0.95\textwidth]{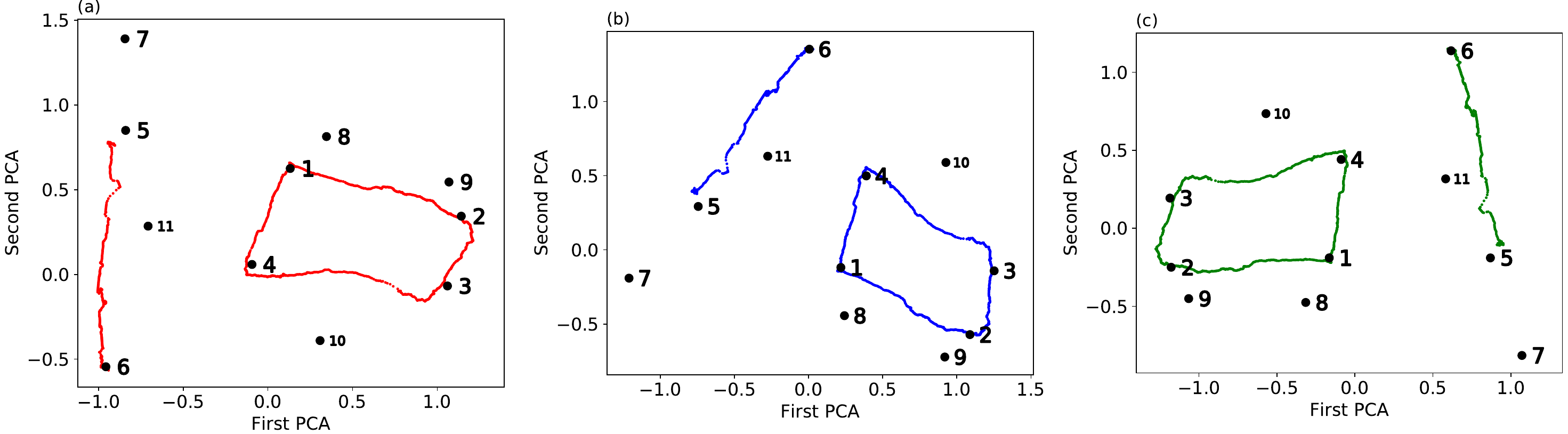}
\caption{PCA method projections, $\bf a$, \eqref{eq:pca} for sliding source along  table edge (points 1-2-3-4). The singular vectors are determined based on  (a) table (point 1--6), (b) chair (point 7--11), and (c) chair+table (point 1--11). The PCA of the mean DOA for the 11 calibration points are also shown. Since the singular vectors differ for the three set of measurements, their mapping is different.  }
\label{fig:PCAmap}
\end{figure*}
\begin{figure*} % FIGURE 8
\centering
\includegraphics[width=0.95\textwidth]{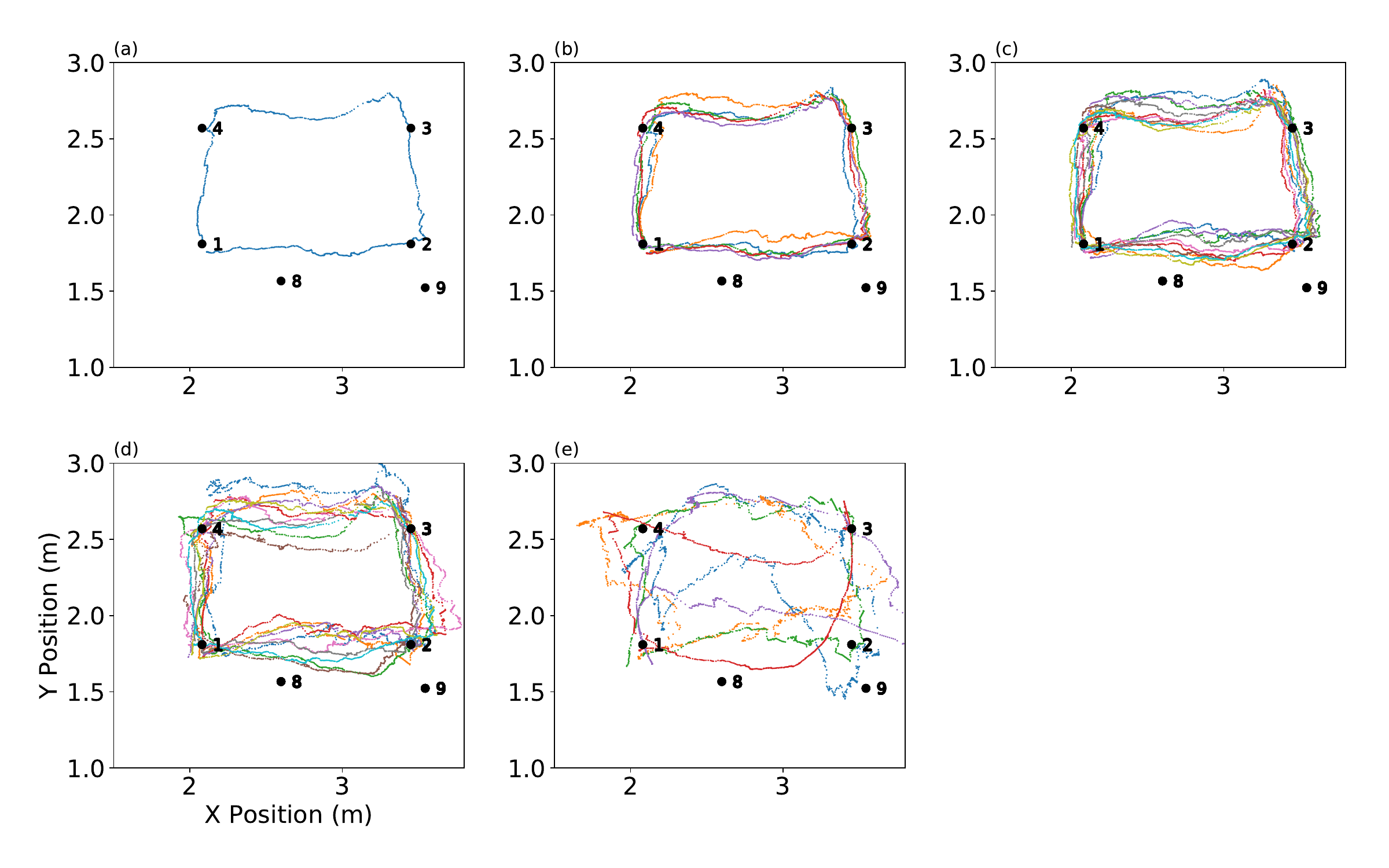}
\caption{Mapping \eqref{eq:Nonlin} based on sliding source along the rectangle of table edge (points 1-2-3-4) with (a) All 5 arrays working or (b) 1, (c) 2, (d) 3 (using 2 arrays), (e) 4 arrays missing (using 1 array). In (b--e) all combinations of  arrays are considered.
}
\label{fig:PCAmapMiss}
\end{figure*}

\subsection{Mapping}
We slide the loudspeaker along the table edge (Fig.~\ref{fig:DOA}) forming ideally  a rectangle (point 1--4) or a line (point 5--6) to demonstrate the method, see Fig.~\ref{fig:maptable}. 
First, a $\bf B$-matrix computed using 2 calibration points was used and we confirmed that in such a case, the mapping was just a projection in the direction of these two points (not shown).
All three  $\bf B$-matrices determined in Fig.~\ref{fig:bmatrix} were used.
 The $\bf B$ matrix based on chair points (blue line)  gives the largest error by drifting away from initial point 1 and 6.
 This is because chair $\bf B$ matrix is determined at different  height and larger horizontal spread,
However, for all mappings it is easy to recognize the rectangle and the line.

\subsection{PCA mapping}
In an actual room setup, it might be impractical to measure the calibration source locations. We could potentially  learn a room from ambient noise DOAs,  and then do a PCA mapping to the first $J=2$ components. For comparison here we use the DOAs from the calibration points as illustrated in Fig.~\ref{fig:PCAmap}. The singular vectors are different for the 3 sets of observation points, see Fig.\ \ref{fig:svd}, but in each of the  PCA projections the shape of the rectangular table and table edge are recognized. In Fig.~\ref{fig:PCAmap},  the PCA component changes due to  changes in the magnitude of the singular vectors entries but the major change is due to sign change of the singular vectors.

The rectangle on the table is 0.76 m $\times$ 1.37 m. Thus, based on Fig.~\ref{fig:PCAmap}c, the long side of the table is ~1.1 units, a change of PCA coordinate, corresponds roughly to a spatial change of 1.37/1.1=1.25 m.
\begin{figure} % FIGURE 9
\centering
	\includegraphics[width=1\columnwidth]{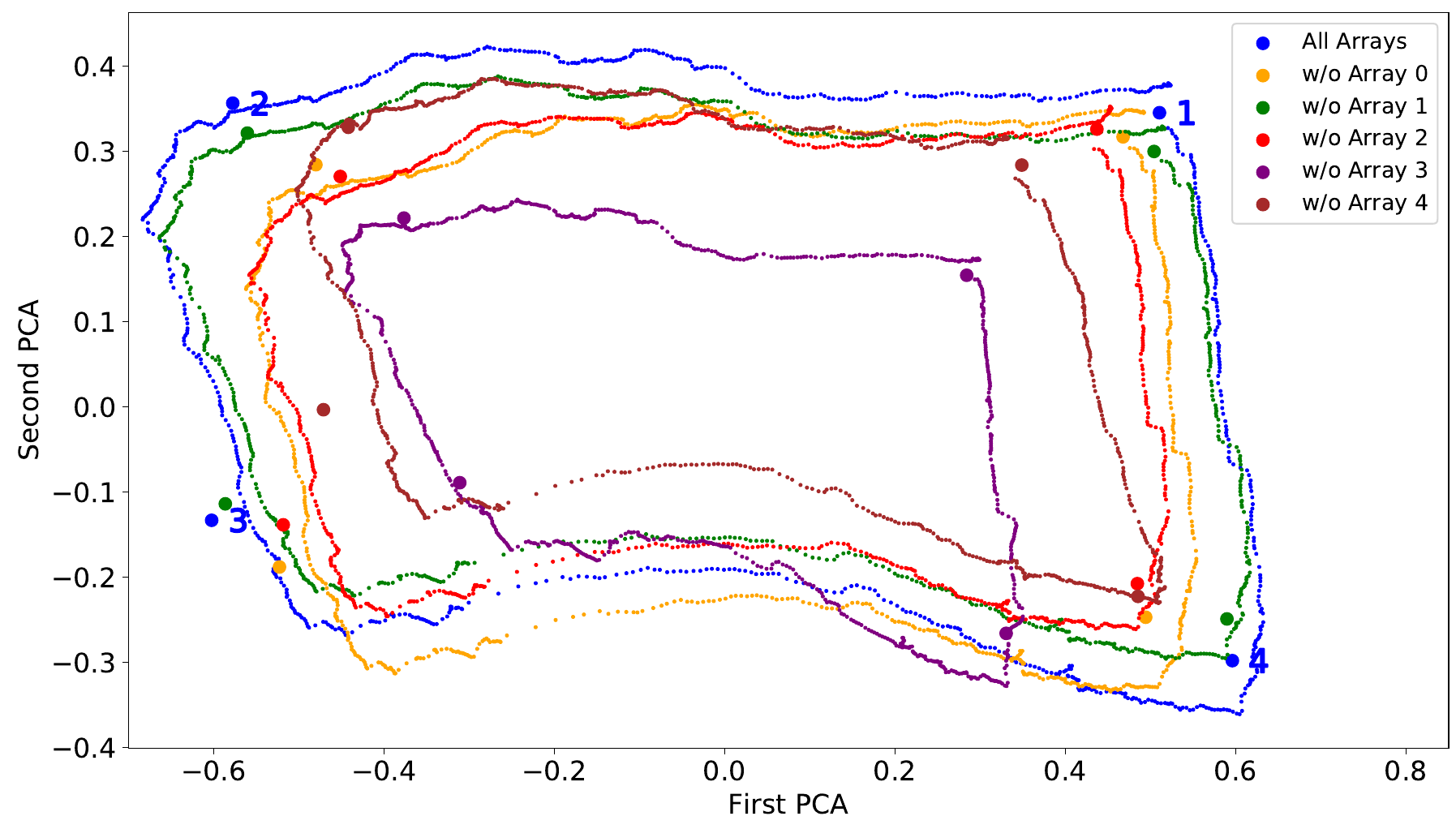}
\caption{PCA for  sliding source along  the  table edge (points 1-2-3-4) with all arrays (blue) or with one array dropped for the whole experiment giving one rectangle for each combination of 4 arrays.
%(a) Full array for whole rectangle and one array missing between calibration point 1 and 2. 
%(b) Mapping for full rectangle. 
When one array is dropped the DOAs for that array is set to 0 and the PCA for the full array is used. It is seen that when one array is dropped the rectangle  maps to other region in PCA space, but can still be identified.
}
\label{fig:mapMiss}
\end{figure}

\begin{figure} % FIGURE 9
\centering
	\includegraphics[width=1\columnwidth]{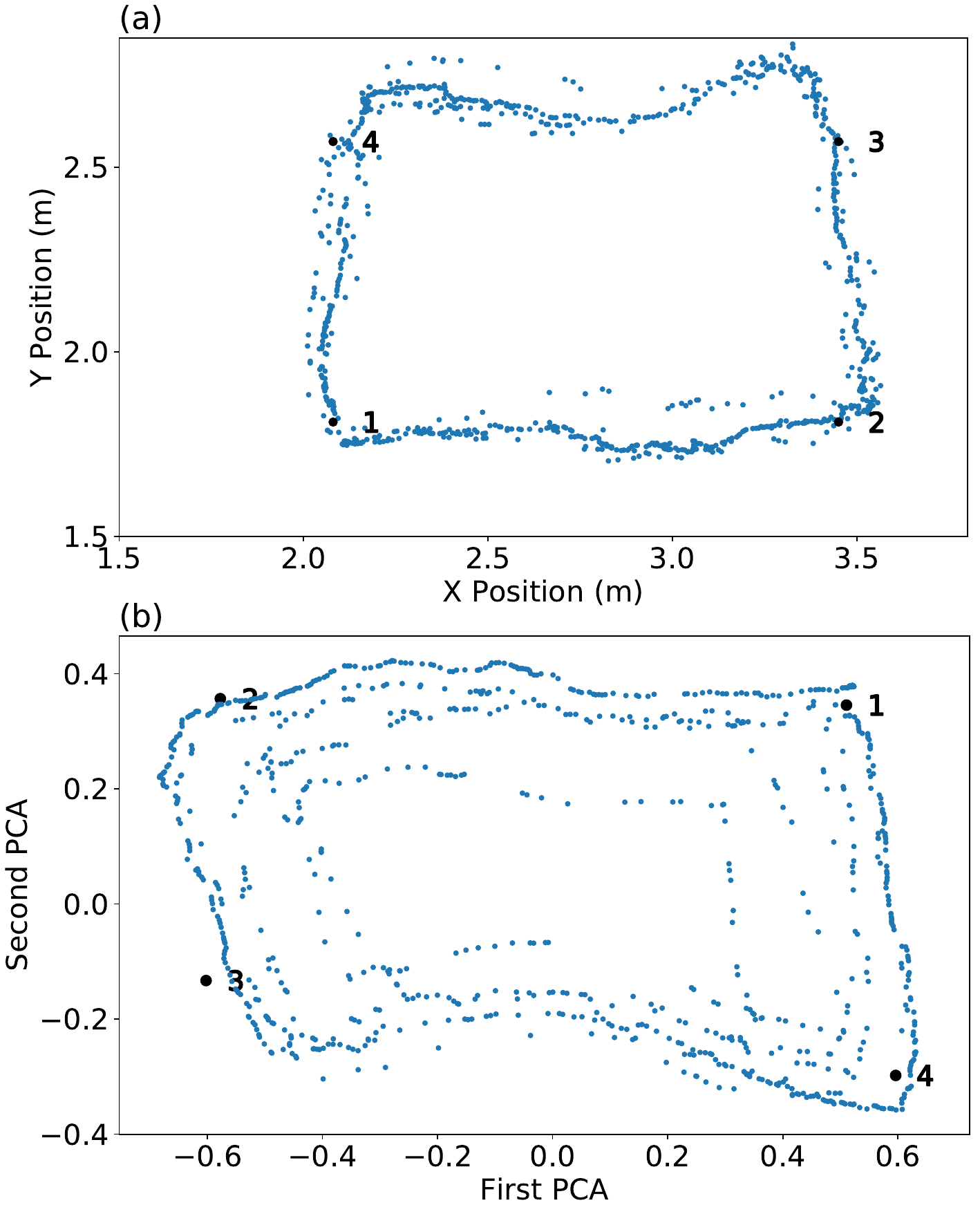}
\caption{Mapping for  sliding source along  the  table edge  (points 1-2-3-4) with for each observation one of the five arrays dropping with 50\% chance.
(a) Mapping \eqref{eq:Nonlin}, and
(b) simple PCA. When mapping to room locations 9a) it is feasible to see the location of the rectangle.
}
\label{fig:mapMissRandom}
\end{figure}

\subsection{Missing arrays}
A weak source can easily cause an array to miss a source, in this subsection we assume that all sources for calibration are sufficiently strong. During the mapping step, the setup could be such that a sound source does not activate the DOA localization for all arrays. This will give an undefined DOA vector for that array. In this section the ${\bf B}$-matrix and PCA are determined by the calibration points on the table (points 1--4).

For the PCA mapping, it is problematic to drop an array as either the PCA has to be recalculated without the array or the missing DOA from that array should be estimated. However, a simple solution is to omit that part of the DOA vector, but still use the same SVD vector based on all arrays. Fig.~\ref{fig:PCAmapMiss} shows how this affects the estimation for the whole rectangle. Depending on which array is dropped it projects to quite different area of the PCA. Thus, if an array is dropped, it does not give clear results for one observation.  However, Fig.~\ref{fig:PCAmapMiss} shows that if we drop an entire array in computing the SVD with the calibration points and for the PCA, the metasurface shows a consistent result with the rectangle of sources clearly mapped.

The non-linear mapping \eqref{eq:Nonlin} maps to the domain and therefore shows good results as long as there are sufficient arrays to perform the mapping, see 
Fig.~\ref{fig:mapMiss}.  Since all calibration points and measurements are in a horizontal plane, even just one array (Fig.~\ref{fig:mapMiss}e) can do well in the mapping provided its DOA are nearly perpendicular to the plane of motion of the sound source. Here, the ceiling arrays (array-2, array-3, array-4) have their DOAs approximately perpendicular to the plane of the motion of the sound source. For one array, the arrays on the walls (array-0, array-1) have DOAs to close to parallel with the source plane, and hence do not give good result. There is significant stability improvement in increasing to two arrays (Fig.~\ref{fig:mapMiss}d).

Figure \ref{fig:mapMissRandom} illustrate the difference in (a) affine mapping \eqref{eq:Nonlin} and (b) PCA mapping for each observation  50\% chance one array is dropped, which array is dropped is decided uniformly. In this figure the ${\bf B}$-matrix is determined by the calibration points on the table (points 1--4). Similar to the above results, the figure shows that simple PCA  is noisy with large uncertainty.

\section{Conclusion}
Sound source localization in rooms with redundant un-localized arrays are discussed.
Each array performs the direction of arrival (DOA) estimation independently and feed it to a cluster where the DOA from all arrays are concatenated. The DOAs can then be used for source localization, either in principal component space or directly mapping to the room.

The method was demonstrated with five circular microphone arrays at un-measured locations in an office. Both methods work well for tracking a source in a room. The direct mapping to room coordinates is more robust to array failures. The methods demonstrated here provides a step towards monitoring activities in smart home with little installation effort.

\section*{Acknowledgements}
We appreciate initial assistance of  Brian Wang and Yifan Wu. 

\bibliographystyle{unsrt}
%\section*{References}
\bibliography{odas,PhaseREFOct2019,biblio_gen_DOA0}% 
\end{document}